\documentclass[aps,pre,twocolumn,superscriptaddress,floatfix,nofootinbib,longbibliography]{revtex4-2}
\usepackage{amsmath,amssymb,graphicx,xcolor,booktabs}
\usepackage{tikz}
\usetikzlibrary{arrows.meta,calc,decorations.pathmorphing}
\usepackage[colorlinks=true,linkcolor=blue!50!black,citecolor=blue!50!black,urlcolor=blue!50!black]{hyperref}
\newcommand{\OmH}{\Omega_H}
\begin{document}
\title{Engineering a digital acoustic black hole}
\author{Antonino Flachi}
\affiliation{Department of Physics \& Research and Education Center for Natural Sciences, Keio University, Yokohama, Japan}
\date{\today}
\begin{abstract}
We show how to design and build an analogue black hole as a synchronous digital circuit. We define the sound speed, the flow and the horizon of such a circuit in terms of its own logic, and realise one on a field-programmable gate array, where the black hole is probed electrically with a signal generator and an oscilloscope. We measure the stability of the horizon, its causal cone,
the ringdown and rotational superradiance. A single flipped bit inside the horizon leaks out by an amount that falls exponentially with depth. Digital circuits thus form a reconfigurable class of gravitational analogues and realise an analogue black hole as a cellular automaton.
\end{abstract}
\maketitle

\section*{Introduction}

In 1981, Unruh identified an analogy between black holes and flowing fluids~\cite{Unruh1981}: sound cannot travel upstream in a fluid that moves faster than sound, just as light cannot escape from a black hole. Sound waves in a moving fluid obey the same equation as a field propagating in curved spacetime, with the surface where the inward flow reaches the speed of sound acting as an event horizon.

Such analogue black holes bring into the laboratory phenomena like Hawking radiation~\cite{Hawking1974}, the ringing of a perturbed black hole, and the amplification of waves by a rotating one (superradiance~\cite{Zeldovich1971,Brito2015}), turning theoretical questions into experimental ones: whether these effects survive the short-distance structure of the medium; how a horizon acts on the waves that reach it; and which features of black holes are universal, and which are particular to Einstein's equations. Laboratory experiments have realized analogue black holes with flowing water~\cite{Weinfurtner2011,Euve2016,Torres2017,Torres2020}, ultracold atomic gases~\cite{Steinhauer2016,deNova2019}, optical fibres~\cite{Philbin2008,Drori2019}, superfluid helium~\cite{Svancara2024} and superconducting circuits, proposed as Hawking-radiation analogues~\cite{Schutzhold2005,Nation2009} and realised as a qubit chain~\cite{Shi2023}. Early theoretical work includes Refs.~\cite{Jacobson1991,Unruh1995,Visser1998,Garay2000,FischerVisser2002,Schutzhold2002}; See Ref.~\cite{BarceloLR} for a review.

Here we describe and realise a new class of analogue black holes, implemented as synchronous digital circuits on a field-programmable gate array chip. The chip's registers hold a grid of cells, and a local update rule, wired as logic gates, is applied to every cell at each clock tick. Specific choices of the update rule produce a flow with a horizon. Such a device is an ordinary digital circuit and can be probed by injecting an electrical signal, which is converted into a wave that scatters off the horizon; the response is read as a voltage and recorded on an oscilloscope.
It is a classical digital device: its horizon is a property of the state of a synchronous logic circuit, i.e. a cellular automaton in which every quantity is a word of bits. Each cell of the digital circuit stores a finite number of bits in its registers, so the wave, the flow and the horizon take only finitely many values, and the dynamics is an exact map on this finite set of states. What such a device offers is an analogue black hole whose short-distance structure is known exactly. Its dispersion relation follows from the update rule of the circuit, and whether horizon physics survives this structure can be tested by direct measurement.

The device is used to measure five properties of the analogue black hole, each read as a voltage: its stability, the horizon itself, the causal cone that defines it, the ringdown (the quasinormal modes of the hole in its reservoir) and superradiance.

\section*{Synchronous digital circuits as analogue spacetimes}

The defining element of black hole analogues is the emergence of a horizon; this requires a disturbance, a speed at which the disturbance spreads, and a flow that carries it. In a fluid these are sound, the speed of sound and the velocity of the fluid. Each has a precise counterpart in a synchronous digital circuit.

A synchronous digital circuit consists of registers, which hold bits, combinational logic, which computes new bits from old ones, and a clock: at every tick all registers are replaced at once by the output of the logic~\cite{HarrisHarris2013}. Let the registers form a regular array of sites, each holding a word $s_x$ of $n$ bits, and let the logic, the same at every site, read only the words within a distance $R$~\cite{ToffoliMargolus1987}:
\begin{equation}
s_x(t+1)=\Phi\big(\{s_{x+d}(t)\}_{|d|\le R};\,b_x,\,a(t)\big),
\label{eq:sdc}\end{equation}
with $\Phi$ a fixed Boolean map, $b_x$ static bits that may differ from site to site, $a(t)$ an external input, and $|d|=\max(|d_x|,|d_y|)$ so that the neighbourhood is a square. 

\begin{figure*}[t]\centering
\resizebox{\textwidth}{!}{%
\begin{tikzpicture}[font=\sffamily\scriptsize,>={Stealth[length=1.5mm]},
  box/.style={draw,rounded corners=1.5pt,align=center,inner sep=2.5pt,fill=white},
  wire/.style={->,thick}]
\begin{scope}[shift={(0,0)}]
\node[font=\sffamily\bfseries\small,anchor=west] at (-0.3,3.1) {a};
\node[box,minimum width=2.6cm,minimum height=1.5cm,fill=gray!8] (reg) at (1.1,1.6) {};
\node[anchor=south] at (reg.north) {registers: the words $s_x$, $n$ bits each};
\foreach \i in {0,...,7} {\pgfmathsetmacro{\b}{int(mod(\i*7+3,5)<2)} \node[draw,fill=white,minimum size=0.26cm,inner sep=0,font=\sffamily\tiny] at (0.15+0.27*\i,1.85) {\b};}
\foreach \i in {0,...,7} {\pgfmathsetmacro{\b}{int(mod(\i*3+1,4)<2)} \node[draw,fill=white,minimum size=0.26cm,inner sep=0,font=\sffamily\tiny] at (0.15+0.27*\i,1.5) {\b};}
\node[font=\sffamily\tiny] at (1.1,1.15) {state at tick $t$};
\node[box,minimum width=2.6cm,minimum height=1.5cm,fill=green!10] (log) at (5.4,1.6) {$\Phi$\\[1pt]combinational logic,\\the same at every site};
\draw[wire] (reg.east) -- node[above,font=\sffamily\tiny]{read} node[below,font=\sffamily\tiny]{$\{s_{x+d}(t)\}_{|d|\le R}$} (log.west);
\draw[wire] ($(log.north)+(-0.8,0.6)$) node[above,font=\sffamily\tiny,align=center]{$b_x$\\static bits} -- ($(log.north)+(-0.8,0)$);
\draw[wire] ($(log.north)+(0.8,0.6)$) node[above,font=\sffamily\tiny,align=center]{$a(t)$\\input} -- ($(log.north)+(0.8,0)$);
\draw[wire] (log.east) -- ++(0.5,0) |- ($(reg.south)+(0,-0.55)$) -| node[pos=0.25,below]{write back $s_x(t+1)$, all at once} (reg.south);
\node[anchor=north west,align=left,text width=6.8cm,font=\sffamily\tiny] at (-0.3,-1.95) {$b_x$ may differ from site to site and lays out the medium; $a(t)$ is the only input; everything else is the state.};
\draw[->] (0.0,-1.45) -- (6.0,-1.45) node[right]{$t$};
\draw[thick] (0.1,-1.3) \foreach \i in {0,...,3} { -- ++(0.7,0) -- ++(0,0.35) -- ++(0.7,0) -- ++(0,-0.35)};
\foreach \i in {0,...,3} {\draw[->,gray] (0.8+1.4*\i,-0.9) -- (0.8+1.4*\i,-0.6);}
\node[anchor=west] at (0.0,-0.45) {clock: at every edge, every register takes the output of the logic};
\node[anchor=east,font=\sffamily\tiny] at (0.0,-1.1) {clock};
\node[anchor=west,font=\sffamily\tiny] at (0.5,-1.75) {tick $t$}; \node[anchor=west,font=\sffamily\tiny] at (1.9,-1.75) {$t+1$}; \node[anchor=west,font=\sffamily\tiny] at (3.3,-1.75) {$t+2$};
\end{scope}
\node[font=\sffamily\bfseries\small,anchor=west] at (8.4,3.1) {b};
\begin{scope}[shift={(9.0,-0.9)}]
 \def\c{0.36}
 \fill[blue!10] (2*\c,2*\c) rectangle (7*\c,7*\c);
 \foreach \i in {0,...,8} \foreach \j in {0,...,8} {\draw[gray!70] (\i*\c,\j*\c) rectangle ++(\c,\c);}
 \fill[blue!60!black] (4*\c,4*\c) rectangle ++(\c,\c);
 \node[white,font=\sffamily\tiny] at (4.5*\c,4.5*\c) {$x$};
 \draw[<->,blue!60!black] (4.5*\c,7.15*\c) -- node[right,font=\sffamily\tiny]{$R$} (4.5*\c,9.2*\c);
 \draw[blue!60!black,thick] (2*\c,2*\c) rectangle (7*\c,7*\c);
 \node[anchor=north,align=center] at (4.5*\c,-0.1) {array of sites; each holds a word $s_x$ of $n$ bits\\ shaded: the words within $|d|\le R$ ($R=2$)};
\end{scope}
\node[font=\sffamily\bfseries\small,anchor=west] at (13.3,3.1) {c};
\begin{scope}[shift={(13.8,-0.9)}]
 \def\c{0.19}
 \foreach \t/\x in {0/0,1/2.6,2/5.2} {
  \begin{scope}[shift={(\x,0)}]
   \foreach \i in {0,...,10} \foreach \j in {0,...,10} {\draw[gray!50,line width=0.2pt] (\i*\c,\j*\c) rectangle ++(\c,\c);}
   \pgfmathsetmacro{\w}{2*\t}
   \fill[black!75] ({(5-\w)*\c},{(5-\w)*\c}) rectangle ({(6+\w)*\c},{(6+\w)*\c});
   \node[anchor=north] at (5.5*\c,-0.1) {$t=\t$};
  \end{scope}}
 \node[anchor=south west,align=left] at (0,11*\c+0.15) {two copies of the circuit, same input, one bit flipped at $t=0$:\\ black, the sites whose words differ};
 \node[anchor=north,align=center] at (2.6+5.5*\c,-0.5) {the difference reaches at most $R$ sites per tick, in every direction};
\end{scope}
\end{tikzpicture}
}
\caption{A synchronous digital circuit as a medium. (a)~Registers hold the words $s_x$; the logic $\Phi$, the same at every site, computes $s_x(t+1)$ from the words within $|d|\le R$, the static bits $b_x$ and the input $a(t)$, Eq.~(\ref{eq:sdc}), and at every clock edge all registers are replaced at once. (b)~The registers as an array of sites, each holding a word $s_x$ of $n$ bits; shaded, the words within $|d|\le R$ of $x$ (drawn for $R=2$). (c)~Two copies run with the same input and one bit flipped at $t=0$: the sites at which they may differ (black) are those within $Rt$ of the flipped bit, a square of side $2Rt+1$, a cone of slope $R$ in every direction. This is the speed limit of the wiring.}
\label{fig:sdc}\end{figure*}
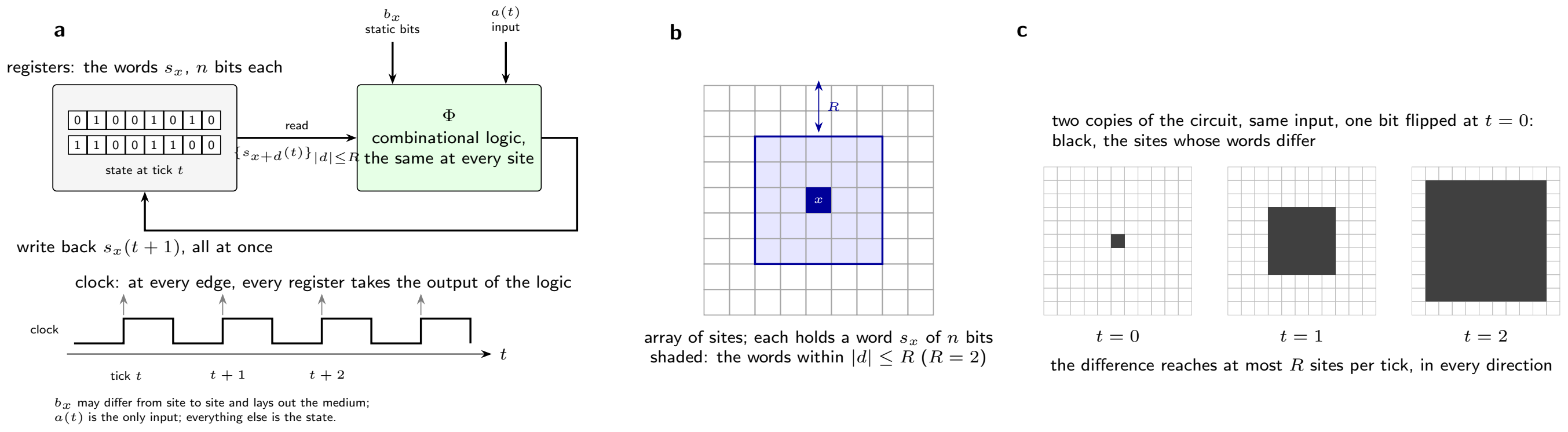

The circuit is deterministic, so a disturbance has an exact meaning. Consider two copies of the circuit, run with the same input, whose states at $t=0$ are identical except for one flipped bit, and follow the sites at which the two copies hold different words. After one tick a word can differ only if $\Phi$ read a word that differed, that is, only within $R$ sites of the flipped bit: at most a $(2R+1)\times(2R+1)$ square, and after $t$ ticks at most a square of side $2Rt+1$. Whether the difference fills that square depends on $\Phi$ and on the state. In the plane of position and time, the sites that may differ fill a cone with apex at the flipped bit and slope $R$: the cone of the wiring (a classical analogue of the Lieb--Robinson bound of lattice quantum systems~\cite{LiebRobinson1972,HastingsKoma2006}). It is the same at every site and in every direction, and by itself it contains no horizon. Figure~\ref{fig:sdc} draws the circuit, the rule and this cone for $R=2$.

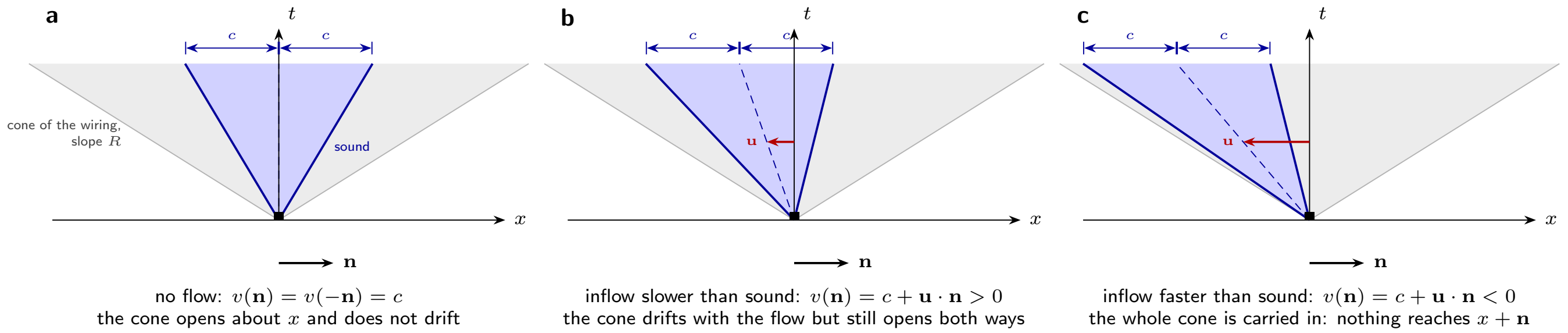
\begin{figure*}[t]\centering
\resizebox{\textwidth}{!}{%
\begin{tikzpicture}[font=\sffamily\scriptsize,>={Stealth[length=1.5mm]},
  cone/.pic={
    \fill[gray!15] (0,0) -- (-\R*2.0,2.0) -- (\R*2.0,2.0) -- cycle;
    \draw[gray!60] (0,0) -- (-\R*2.0,2.0); \draw[gray!60] (0,0) -- (\R*2.0,2.0);
    \fill[blue!18] (0,0) -- ({(\u-\cc)*2.0},2.0) -- ({(\u+\cc)*2.0},2.0) -- cycle;
    \draw[blue!60!black,thick] (0,0) -- ({(\u-\cc)*2.0},2.0);
    \draw[blue!60!black,thick] (0,0) -- ({(\u+\cc)*2.0},2.0);
    \draw[blue!60!black,densely dashed] (0,0) -- ({\u*2.0},2.0);
    \draw[->] (-2.9,0) -- (2.9,0) node[right]{$x$};
    \draw[->] (0,0) -- (0,2.45) node[above right]{$t$};
    \fill[black] (-0.06,0) rectangle (0.06,0.1);
    \draw[->,thick] (0.0,-0.55) -- (0.7,-0.55) node[right]{$\mathbf n$};
    \draw[|<->|,blue!60!black] ({\u*2.0},2.2) -- ({(\u+\cc)*2.0},2.2) node[midway,above,font=\sffamily\tiny]{$c$};
    \draw[|<->|,blue!60!black] ({(\u-\cc)*2.0},2.2) -- ({\u*2.0},2.2) node[midway,above,font=\sffamily\tiny]{$c$};
  }]
\def\R{1.6}
\begin{scope}[shift={(0,0)}]
 \def\u{0} \def\cc{0.6}
 \node[font=\sffamily\bfseries\small] at (-2.9,2.6) {a};
 \pic {cone};
 \node[anchor=north,align=center] at (0,-0.75) {no flow: $v(\mathbf n)=v(-\mathbf n)=c$\\ the cone opens about $x$ and does not drift};
 \node[gray!50!black,font=\sffamily\tiny,anchor=east,align=right] at (-1.9,1.1) {cone of the wiring,\\ slope $R$};
 \node[blue!60!black,font=\sffamily\tiny,anchor=west] at (0.6,0.95) {sound};
\end{scope}
\begin{scope}[shift={(6.6,0)}]
 \def\u{-0.35} \def\cc{0.6}
 \node[font=\sffamily\bfseries\small] at (-2.9,2.6) {b};
 \pic {cone};
 \draw[->,red!70!black,thick] (0,1.0) -- ({\u*2.0*0.5},1.0) node[left,font=\sffamily\tiny]{$\mathbf u$};
 \node[anchor=north,align=center] at (0,-0.75) {inflow slower than sound: $v(\mathbf n)=c+\mathbf u\cdot\mathbf n>0$\\ the cone drifts with the flow but still opens both ways};
\end{scope}
\begin{scope}[shift={(13.2,0)}]
 \def\u{-0.85} \def\cc{0.6}
 \node[font=\sffamily\bfseries\small] at (-2.9,2.6) {c};
 \pic {cone};
 \draw[->,red!70!black,thick] (0,1.0) -- ({\u*2.0*0.5},1.0) node[left,font=\sffamily\tiny]{$\mathbf u$};
 \node[anchor=north,align=center] at (0,-0.75) {inflow faster than sound: $v(\mathbf n)=c+\mathbf u\cdot\mathbf n<0$\\ the whole cone is carried in: nothing reaches $x+\mathbf n$};
\end{scope}
\end{tikzpicture}
}
\caption{Sound speed and flow velocity read off the cone. Space--time diagrams of the sites reached by the difference between two copies of the circuit after one bit is flipped at $x$ (black mark): the cone of the wiring (grey, slope $R$) and, inside it, the sound cone (blue), whose edges advance at $v(\mathbf n)$ and $v(-\mathbf n)$; dashed, its centre. (a)~No flow. (b)~Inflow slower than sound: the cone drifts with the flow (red arrow) but still opens on both sides. (c)~Inflow faster than sound: $v(\mathbf n)<0$, the cone is carried inward and no disturbance reaches the outer side; the horizon is where this first happens.}
\label{fig:conedef}\end{figure*}

How the disturbance propagates inside the cone of the wiring depends on $\Phi$ and on the state of the circuit. Suppose that $\Phi$ implements conservation laws. A physical example is a ``mass'' associated with each site, $m_x=\sum_i f_i(x)$ with $s_x=\{f_0(x),f_1(x),\dots\}$ the word read as a list of numbers, and a ``total mass'' $m_{\rm tot}=\sum_x m_x$ that the conservation law requires to stay constant. The same can be done for a quantity analogous to the momentum, $\sum_i \mathbf e_i f_i(x)$ with $\mathbf e_i$ a direction attached to each entry. Such definitions are those of a lattice-Boltzmann scheme~\cite{ToffoliMargolus1987,Succi2001} and allow one to interpret the evolution of the digital circuit, on scales much larger than $R$, as that of a fluid. Assuming the circuit settles to a steady state, a small disturbance of that state travels as a sound wave, at a speed $c$ set by $\Phi$ and by the state and bounded by $R$. The bulk of the difference between the two copies then stays inside a narrower cone, that of the sound; the rest spreads by dispersion at the speed of the wiring, decaying at every tick.

 Let $v(x,\mathbf n)\le R$ be the speed at which the edge of this narrower cone, the one that holds the bulk of the difference made by a bit flipped at $x$, advances in the direction $\mathbf n$; it may be negative. After $t$ ticks the cone extends from $x-v(x,-\mathbf n)\,t$ to $x+v(x,\mathbf n)\,t$ along $\mathbf n$: its width grows at the rate $v(x,\mathbf n)+v(x,-\mathbf n)$ and its centre moves at the rate $v(x,\mathbf n)-v(x,-\mathbf n)$. This suggests to define
\begin{equation}
\begin{aligned}
c(x)&=\tfrac12\big[v(x,\mathbf n)+v(x,-\mathbf n)\big],\\
\mathbf u(x)\cdot\mathbf n&=\tfrac12\big[v(x,\mathbf n)-v(x,-\mathbf n)\big]
\end{aligned}
\label{eq:cu}\end{equation}
as the rate at which the cone opens and the rate at which it drifts: the speed of sound of the circuit and the velocity of its flow, defined by the circuit properties alone. If the rule disperses, $c$ and $\mathbf u$ are then the rates for disturbances long compared with the spacing of the sites, and the shorter ones run ahead or behind as precursors. Figure~\ref{fig:conedef} shows the cone with no flow, with a subsonic and with a supersonic flow.

We can define a horizon as a closed curve $\Sigma$, with outward normal $\mathbf n$, on which $v(x,\mathbf n)=0$, with $v<0$ just inside. On $\Sigma$ the state carries differences inward as fast as they spread, and inside it faster. Where the cone of long disturbances does not contain the site it starts from, $v(x,\mathbf n)<0$ for some direction $\mathbf n$, that is $|\mathbf u|>c$, the circuit has an ergoregion.

The horizon belongs to the state of the circuit, not to its wiring: the wires cross $\Sigma$ in both directions and the logic is the same on both sides.  On scales of a few sites a perturbation contains a dispersive part, which spreads at the speed of the wiring. This part is suppressed at long scales, but at short scales it may cross $\Sigma$, yielding a small leak. If short disturbances travel faster than long ones, as they do in the circuit built below, the leak comes from just inside $\Sigma$ and falls off with depth: the horizon is sharp for long disturbances and leaky for short ones. This is reminiscent of analogue horizons with superluminal dispersion~\cite{Corley1996,Macher2009}.

\section*{Design of a synchronous digital circuit as an analogue black hole}

The construction of the previous section is formal: it does not guarantee that for a given map a horizon forms or that it is stable. These are properties of a running device, to be measured through an input $a(t)$ and an output read from its state. 

The design of the device corresponds to a choice of $s_x$, the map $\Phi$, the static bits $b_x$ and the input $a(t)$. Here, with the shallow-water model of black holes in mind~\cite{Schutzhold2005,Weinfurtner2011}, we choose them so that, on large scales, the circuit mimics a shallow layer of water flowing over a bed. As in shallow water, the wave speed depends on the depth, so the speed of sound of the circuit becomes a property of its state. We follow the construction of ref.~\cite{FlachiSW}. Each word is a set of $25$ floating-point numbers $f_i$ of $32$ bits, one for each of a fixed set of directions $\mathbf e_i$ reaching up to two sites, so that $n=800$; the push reads the depth at the nearest neighbours, so that $\Phi$ reads the words within $R=3$ sites. The map $\Phi$ acts in two steps. First, each site forms the depth $h=\sum_i f_i$ and the momentum $h\mathbf u=\sum_i\mathbf e_i f_i$ adds a push set by the slopes of the depth and of the bed at its neighbours, and relaxes its $f_i$ towards a fixed function of $h$ and $\mathbf u$. Then each $f_i$ moves to the site its direction points to. The relaxation and the moves conserve depth and momentum, up to the rounding of $32$-bit arithmetic; the push only exchanges momentum between neighbouring sites, except for the slope of the bed, which acts as an external force. Water enters or leaves only at the reservoir, the drain, the outflow and the drive described below, and the absorbing layer removes momentum. On $32$-bit words it is a Boolean map of the form of Eq.~(\ref{eq:sdc}), whose long-wavelength limit is the shallow-water fluid, with $c=\sqrt{gh}$ and $\mathbf u$ the velocity of the water.

The static bits $b_x$ lay out the analogue black hole: the reservoir (the sites beyond an outer radius, whose words are reset at every tick to a layer of fixed depth flowing inward with a swirl), the drain (the central sites, whose words are scaled down at every tick so that a fixed volume leaves; it forms the hole and is switched off once the sill is on), the sill (a ring of sites carrying a raised bed, over whose crest the inflow crosses the wave speed, which holds the horizon away from the centre), the outflow (the sites inside the sill, which copy at every tick the state of a ring just outside them, so that the water leaves freely), and the drive ring (the sites that receive the input). The input $a(t)$ consists of two numbers per tick, added to the words of the drive ring with angular patterns $\cos m\theta$ and $\sin m\theta$: they set the amplitude and the sense of rotation of a wave of azimuthal number $m$ sent towards the hole. The controls use the same circuit: with the swirl of the reservoir set to zero the hole does not rotate, and reversing the sign of one input reverses the sense of the wave. The drain and the outflow are not of the form of Eq.~(\ref{eq:sdc}): the drain uses the total depth of the central sites, and the outflow copies sites along a ray. In the held state only the outflow acts, inside the horizon, and its sites are left out of the causal-cone analysis (Appendix).

The design of the circuit as a shallow-water fluid allows one to predict where the horizon forms: where $v(x,\hat{\mathbf r})=c-|u_r|$ crosses zero, that is over the crest of the sill. The swirl makes the flow rotate, at an angular velocity $\OmH$ on $\Sigma$, with an ergoregion just outside.

Treating the circuit as a fluid is only formal. Whether a real circuit behaves as an analogue black hole is a question that only a measurement answers. In the following section we describe how this design is implemented in hardware, where it can be probed by electrical signals.

\section*{Hardware implementation}

The circuit is implemented by translating the map $\Phi$ into a hardware description, registers and logic, placed and routed on a field-programmable gate array~\cite{HarrisHarris2013}.  In the configured chip the map exists as gates and wires, with no instructions in between.

A hardware implementation of Eq.~(\ref{eq:sdc}) would require one copy of $\Phi$ per site, far more than the chip holds. We therefore adopt the standard alternative for array (stencil) computations in hardware, to keep the words of the array in the board's memory and apply a single copy of $\Phi$ to the sites one after another~\cite{Kung1982,Sano2007}. Every site is updated from the words of the previous tick, so the folded circuit computes the same bits as the unfolded one (Fig.~\ref{fig:device}, the Appendix).

The processor on the same chip starts each tick, passes the two input numbers to the drive, forms the ring sums that give the output, and writes it. The state of the circuit evolves in the logic alone, so the processor affects neither the evolution of the circuit nor the nature of the measurement. The device is deterministic, and this allows the circuit to be checked at the three stages of its construction; we find that the chip computes $\Phi$ as expected. Each protocol is given in the Appendix.

\section*{Measurements}

The analogue black hole on the chip is probed by a two-channel signal generator, which supplies the input $a(t)$ through analogue-to-digital converters, and by an oscilloscope, which records the observable written to a digital-to-analogue converter (Fig.~\ref{fig:device}, the Appendix). The figures below show the voltages recorded on the oscilloscope together with the board's digital record of the same quantity and the value of the $64$-bit numerical twin~\cite{FlachiSW}; the uncertainties of the oscilloscope readings are given in the Appendix.

\begin{figure*}[t]\centering
\resizebox{\textwidth}{!}{%
\begin{tikzpicture}[font=\sffamily\scriptsize,>={Stealth[length=1.5mm]},
  box/.style={draw,rounded corners=1.5pt,align=center,inner sep=2.5pt,fill=white},
  pan/.style={draw,rounded corners=2pt,fill=gray!4,minimum width=9.3cm,minimum height=4.2cm},
  wire/.style={->,thick},
  hole/.pic={
    \draw[gray!60,thick] (0,0) circle (1.2);
    \draw[gray!60,densely dotted,thick] (0,0) circle (0.98);
    \draw[very thick] (0,0) circle (0.6);
    \fill[black] (0,0) circle (0.08);
    \draw[->,gray!60] (1.2,0) -- (0.64,0); \draw[->,gray!60] (-1.2,0) -- (-0.64,0);
    \draw[->,gray!60] (0,1.2) -- (0,0.64); \draw[->,gray!60] (0,-1.2) -- (0,-0.64);
  }]
\node[box,minimum width=2.1cm] (gen) at (0,0) {generator\\W1 \quad W2};
\node[box,fill=blue!6] (adc) at (3.0,0) {ADC};
\node[box,fill=green!10,minimum width=5.2cm,minimum height=0.9cm] (brd) at (7.0,0) {board: circuit $\Phi$ (logic) + processor};
\node[box,fill=blue!6] (dac) at (11.0,0) {DAC};
\node[box,minimum width=2.1cm] (osc) at (14.0,0) {oscilloscope};
\draw[wire] (gen) -- node[above,font=\sffamily\tiny]{$a(t)$} (adc); \draw[wire] (adc) -- (brd);
\draw[wire] (brd) -- (dac); \draw[wire] (dac) -- node[above,font=\sffamily\tiny]{output} (osc);
\draw[wire,gray] (gen.south) -- ++(0,-0.35) -| node[pos=0.25,below]{drive, for reference} (osc.south);
\def\panel#1#2#3{\node[pan] (#1) at (#2) {}; \node[anchor=north west,font=\sffamily\bfseries\small] at (#1.north west) {#3};}
\panel{A}{2.9,-3.85}{1 \ Stability}
\panel{B}{12.4,-3.85}{2 \ Horizon}
\panel{C}{2.9,-8.45}{3 \ Ringdown}
\panel{D}{12.4,-8.45}{4 \ Superradiance}
\begin{scope}[shift={(2.9,-3.85)}]
 \node[align=center,text width=2.1cm] at (-2.9,0.3) {no input};
 \node[align=center,text width=2.1cm,gray] at (-2.9,-0.5) {the hole forms from rest and is left alone};
 \pic at (-0.2,-0.1) {hole};
 \draw[<->,blue!60!black,thick] (-0.2,-0.1) -- ($(-0.2,-0.1)+(40:0.6)$) node[midway,right=1pt,yshift=-3pt]{$r_h$};
 \node[anchor=south,align=center] at (-0.2,1.2) {state read back every 200 ticks;\\ horizon radius $r_h$ to the DAC};
 \draw[wire] (1.15,-0.1) -- (1.75,-0.1);
 \draw[->] (1.9,-1.1) -- (3.9,-1.1) node[right]{$t$}; \node[below] at (2.9,-1.1) {hours};
 \draw[->] (1.9,-1.1) -- (1.9,0.8) node[above]{$r_h$};
 \draw[thick] (1.95,-0.95) .. controls (2.3,-0.4) and (2.5,0.15) .. (2.8,0.2) -- (3.8,0.2);
 \node[anchor=west,align=left] at (2.0,0.65) {forms, settles, holds};
\end{scope}
\begin{scope}[shift={(12.4,-3.85)}]
 \node[align=center,text width=2.1cm] at (-2.9,0.5) {W1: a probe voltage $V$ sets a radius $r$};
 \draw[->] (-3.7,-1.0) -- (-2.1,-1.0) node[right]{$t$}; \draw[->] (-3.7,-1.0) -- (-3.7,-0.1) node[left]{$V$};
 \draw[thick] (-3.65,-0.9) -- (-2.2,-0.25); \node[anchor=west] at (-3.5,-0.25) {sweep};
 \pic at (-0.2,-0.1) {hole};
 \draw[blue!60!black,thick,densely dashed] (-0.2,-0.1) circle (0.82);
 \fill[blue!60!black] (-0.2,0.72) circle (0.06); \node[blue!60!black,anchor=south west] at (-0.15,0.7) {$r$};
 \node[anchor=south,align=center] at (-0.2,1.2) {returns the ratio $|u_r|/c$\\ of inflow speed to wave speed at $r$};
 \draw[wire] (1.15,-0.1) -- (1.75,-0.1);
 \draw[->] (1.9,-1.1) -- (3.9,-1.1) node[right]{$r$}; \node[below] at (2.9,-1.1) {$\propto V$};
 \draw[->] (1.9,-1.1) -- (1.9,0.8) node[above]{$|u_r|/c$};
 \draw[densely dashed] (1.9,0.0) -- (3.9,0.0) node[right,font=\sffamily\tiny]{1};
 \draw[thick] (1.95,0.65) .. controls (2.5,0.4) and (2.7,0.2) .. (2.8,0.0) .. controls (3.0,-0.4) and (3.4,-0.8) .. (3.85,-0.9);
 \draw[dotted] (2.8,0.0) -- (2.8,-1.1); \node[font=\sffamily\tiny,anchor=south west] at (2.8,0.02) {$r_h$};
\end{scope}
\begin{scope}[shift={(2.9,-8.45)}]
 \node[align=center,text width=2.1cm] at (-2.9,0.5) {W1: one voltage pulse perturbs the flow with pattern $\cos m\theta$};
 \draw[->] (-3.7,-1.0) -- (-2.1,-1.0) node[right]{$t$}; \draw[->] (-3.7,-1.0) -- (-3.7,-0.1);
 \draw[thick] (-3.65,-0.9) -- (-3.1,-0.9) -- (-3.1,-0.3) -- (-2.9,-0.3) -- (-2.9,-0.9) -- (-2.2,-0.9);
 \pic at (-0.2,-0.1) {hole};
 \foreach \a in {0,60,...,300} {\fill[blue!60!black] ($(-0.2,-0.1)+(\a:0.8)$) circle (0.045);}
 \draw[blue!60!black,densely dashed] (-0.2,-0.1) circle (0.8);
 \node[anchor=south,align=center] at (-0.2,1.2) {returns, tick by tick, the multipole $z_m$\\ on a ring outside $\Sigma$};
 \draw[wire] (1.15,-0.1) -- (1.75,-0.1);
 \draw[->] (1.9,-1.1) -- (3.9,-1.1) node[right]{$t$}; \node[below] at (2.9,-1.1) {ticks};
 \draw[->] (1.9,-1.1) -- (1.9,0.8) node[above]{$z_m$};
 \draw[thick,domain=0:1.9,samples=80] plot ({1.95+\x},{-0.4+0.6*exp(-1.4*\x)*cos(deg(11*\x))});
 \node[anchor=west,align=left] at (2.0,0.65) {decaying tones:};
 \node[anchor=west,align=left] at (2.0,0.42) {quasinormal modes};
\end{scope}
\begin{scope}[shift={(12.4,-8.45)}]
 \node[align=center,text width=2.1cm] at (-2.9,0.55) {W1, W2 in quadrature: a rotating wave of chosen sense and frequency $\omega$};
 \draw[->] (-3.7,-1.0) -- (-2.1,-1.0) node[right]{$t$};
 \draw[thick,domain=0:1.5,samples=60] plot ({-3.65+\x},{-0.55+0.3*sin(deg(6*\x))});
 \draw[thick,gray,domain=0:1.5,samples=60] plot ({-3.65+\x},{-0.55+0.3*cos(deg(6*\x))});
 \pic at (-0.2,-0.1) {hole};
 \draw[->,blue!60!black,thick] ($(-0.2,-0.1)+(15:0.98)$) arc (15:75:0.98);
 \node[blue!60!black,font=\sffamily\tiny] at ($(-0.2,-0.1)+(45:1.3)$) {$\pm m$};
 \draw[blue!60!black,densely dashed] (-0.2,-0.1) circle (0.7);
 \node[anchor=south,align=center] at (-0.2,1.2) {returns the net energy flux\\ just outside $\Sigma$};
 \draw[wire] (1.15,-0.1) -- (1.75,-0.1);
 \draw[->] (1.9,-1.1) -- (3.9,-1.1) node[right]{$t$}; \node[below] at (2.9,-1.1) {ticks};
 \draw[->] (1.9,-1.1) -- (1.9,0.8) node[above]{flux};
 \draw[densely dashed] (1.9,-0.3) -- (3.9,-0.3);
 \draw[thick] (1.95,-0.3) \foreach \i in {0,...,9} { -- ++(0,{0.32*(1-2*mod(\i,2))}) -- ++(0.19,0) -- ++(0,{-0.32*(1-2*mod(\i,2))}) };
 \node[anchor=west,align=left,font=\sffamily\tiny] at (2.0,0.55) {sign alternates each tick,};
 \node[anchor=west,align=left,font=\sffamily\tiny] at (2.0,0.35) {offsets cancel on demodulation};
\end{scope}
\end{tikzpicture}
}
\caption{The four measurements, schematically: a signal generator at the input of the chip, an oscilloscope at its output. In each panel the input signal is on the left, the circuit with the horizon $\Sigma$ (thick circle) in the middle and the oscilloscope trace on the right; blue marks where the measurement acts.}
\label{fig:four}\end{figure*}
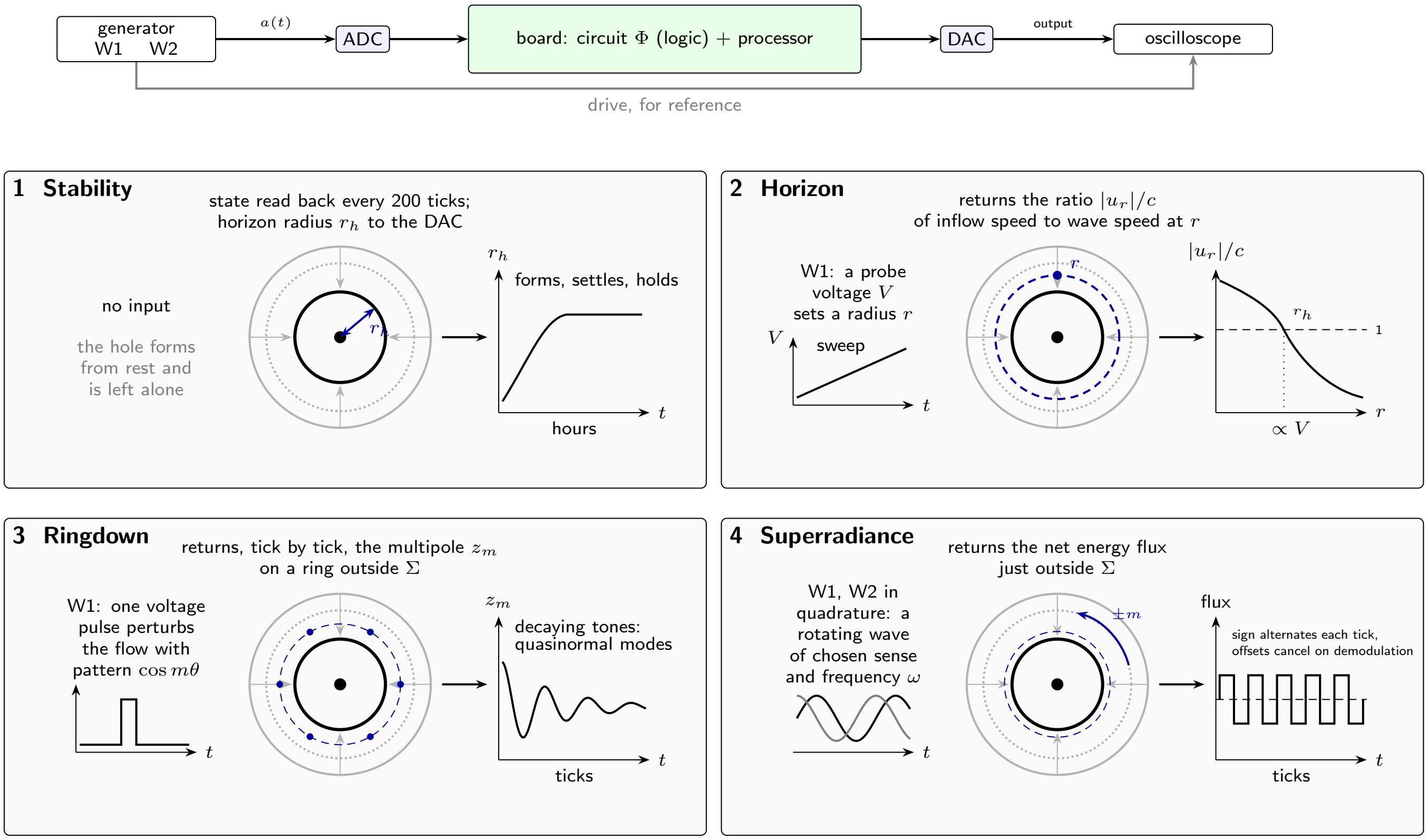

\begin{figure*}[t]\centering
\includegraphics[width=\textwidth]{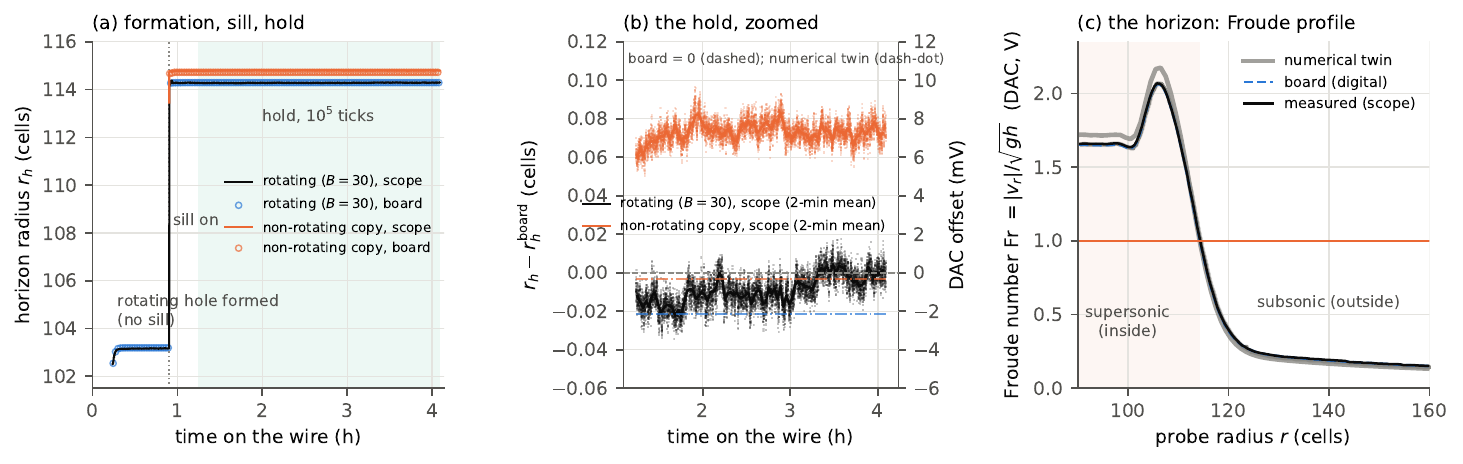}
\caption{(a) Horizon radius against time: formation without the sill, the step when the sill is switched on, and the hold of $1.1\times10^5$ ticks, for the rotating hole and its non-rotating copy; lines, oscilloscope; circles, board record. (b) The hold, zoomed: oscilloscope minus board record; dash-dotted, numerical twin. (c) Froude number, the ratio of inflow speed to wave speed ($1$~V per unit), against the probe radius set by the generator; grey dots, raw samples over twelve sweeps; black, binned; dashed, board record; thick grey, numerical twin.}
\label{fig:hold}\end{figure*}

\emph{Stability.} Every $200$ ticks the board reads back the state of the circuit, finds the radius of the horizon and writes it to the output (Fig.~\ref{fig:four}, panel 1; Fig.~\ref{fig:hold}). Started from rest, the flow settles with its horizon at $103.2$ sites; when the sill is switched on, the horizon moves to $114.29$ sites and stays there to one part in $10^4$ for $1.1\times10^5$ ticks, three hours on the oscilloscope. The non-rotating copy of the hole has its horizon at $114.72$ sites.

\emph{The horizon.} One generator channel sets a probe radius, and the board returns the ratio of the inflow speed to the wave speed at that radius (Fig.~\ref{fig:four}, panel 2). A slow triangular sweep of the generator moves the probe in and out, and the oscilloscope, plotting output against input, draws the profile of the flow (Fig.~\ref{fig:hold}c). The board's record of the same sweeps crosses one at $114.29$, as does the horizon found from the full state.

\begin{figure*}[t]\centering
\includegraphics[width=\textwidth]{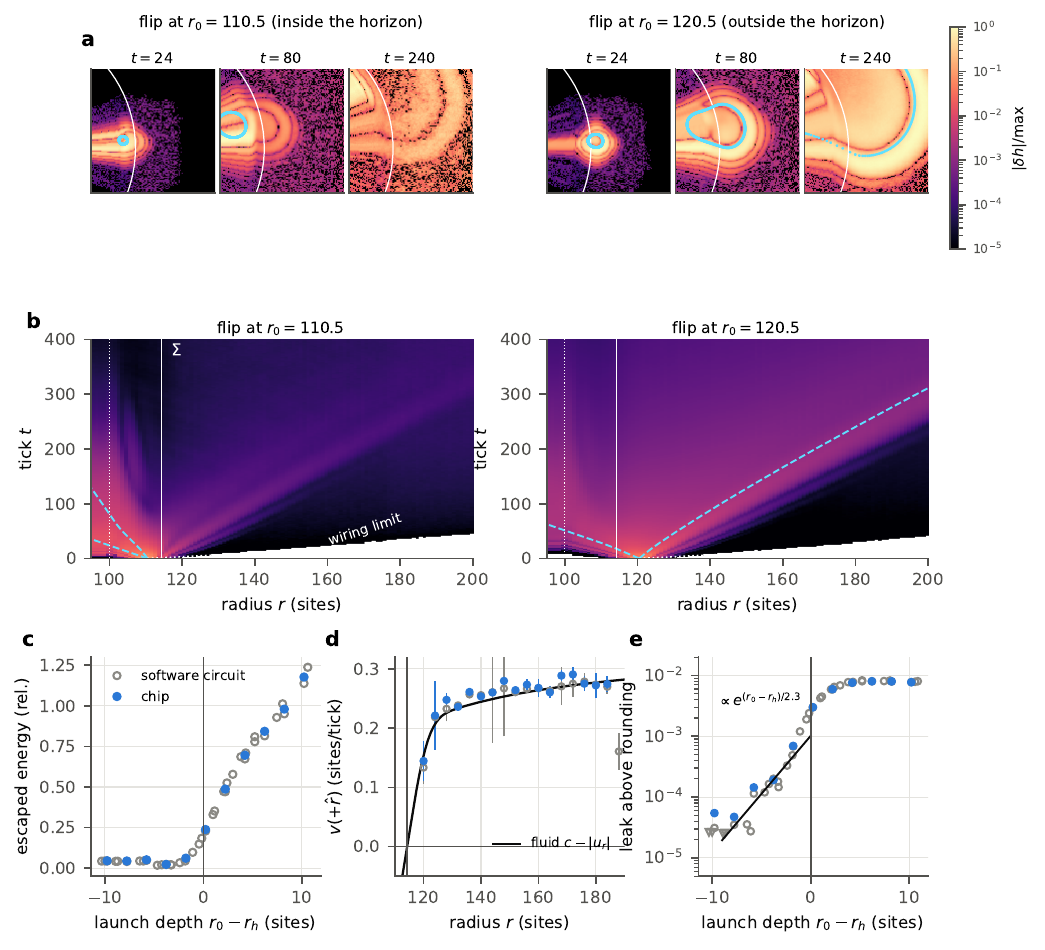}
\caption{(a) Difference of the depth fields of two runs that differ by one bit at $t=0$, flipped inside and outside the horizon, on a logarithmic scale; white circle, the horizon; cyan, the acoustic front of the fluid limit. (b) The same against radius and tick, summed over angle; dashed, the characteristics of the fluid limit; dotted, the wiring limit{ $r_0+3t$}. (c) Energy reaching beyond $135$ sites against the depth of the flip below the horizon, relative to flips well outside; maximum over the run, with the rounding floor included, which sets the level of $0.04$ for deep flips. (d) Outward speed of the difference against radius, with $c-|u_r|$ of the fluid limit. Points, medians over bins of four sites of the centroid speeds measured over $30$ ticks in every run; bars, $1.25$ times the standard error within the bin. (e) Energy beyond $135$ sites at $t=200$ relative to the energy launched, with the rounding floor subtracted; triangles, points within $5\%$ of the floor, drawn as upper limits at that level; filled, chip; open, software circuit; line, exponential fit{ over the three launch angles of the software circuit}.}
\label{fig:cone}\end{figure*}

\emph{The causal cone.} The circuit is run twice from the held (steady) state with one bit flipped between the two runs, and the difference of the depth fields is recorded (Fig.~\ref{fig:cone}). The set of differing words spreads at the speed of the wiring; the bulk of the difference forms a much narrower cone (Fig.~\ref{fig:cone}a,b). When the flipped bit sits outside the horizon, the cone of the difference drifts inward but opens on both sides, as in Fig.~\ref{fig:conedef}b, and its outer edge advances at the speed $c-|u_r|$ (Fig.~\ref{fig:cone}d). When it sits inside, the whole cone is carried inward, as in Fig.~\ref{fig:conedef}c, and its bulk does not reach the outside. The experiment is repeated with the bit flipped at different distances $r_0$ from the centre, and for each run we record the energy of the difference that reaches a circle well outside the horizon. Fig.~\ref{fig:cone}c, d, e show that the horizon is sharp for long waves and leaky for short ones, by an amount that falls exponentially with the depth of the flipped bit inside the horizon. The leak starts as short waves: in what first arrives beyond $135$ sites, $83\%$ of the power of $\delta h$ above the rounding floor lies at $k>0.6$, whereas inside $\Sigma$ $90\%$ lies below it (Fig.~\ref{fig:cone2}b). It also depends on direction, as the dispersion of the rule does (Fig.~\ref{fig:disp}): on the chip the decay length is $3.1\pm0.6$ sites along a lattice axis and $1.9\pm0.2$ along a diagonal. A flipped bit leaks far more than a smooth depth bump, because it excites every wavenumber of the lattice, including the short waves that climb out. The scale of the fall-off and its dependence on direction follow from the dispersion of the rule, with nothing fitted (Appendix).

\begin{figure*}[t]\centering
\includegraphics[width=\textwidth]{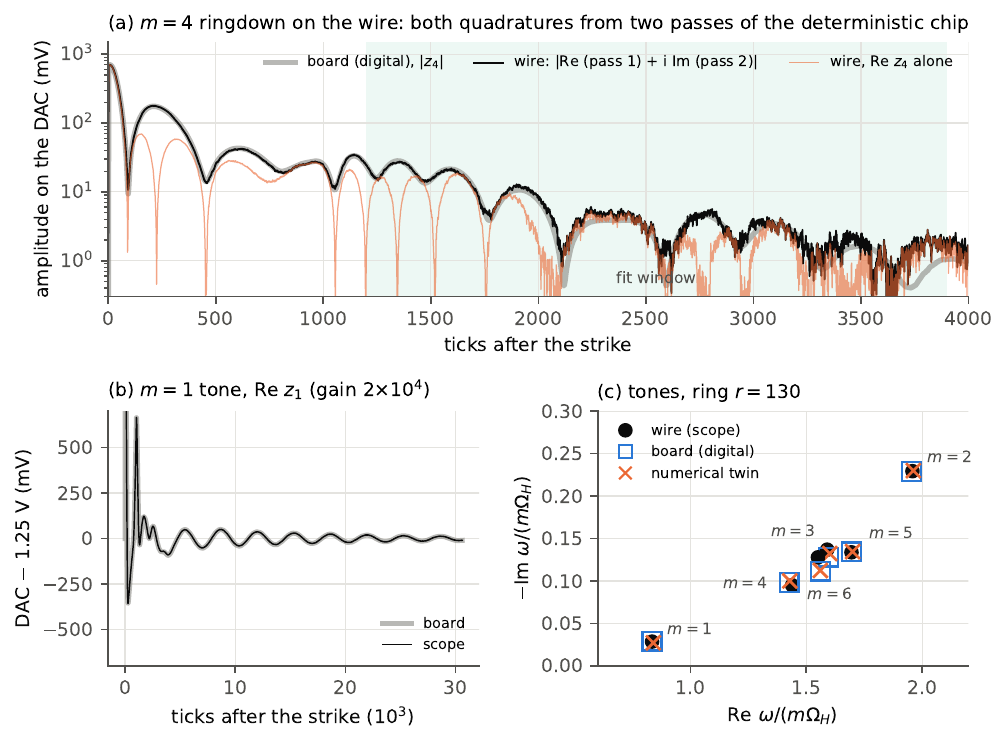}
\caption{Ringdown on the oscilloscope. (a) Amplitude of the $m=4$ response after the pulse, rebuilt from the
two passes (one per component), against the board record; the shaded window is the one used for the tones.
(b) The $m=1$ tone, recorded for $3\times10^4$ ticks. (c) Frequencies and damping rates of the tones, $m=1$ to
$6$, in units of $m\OmH$: oscilloscope, board record and numerical twin.}
\label{fig:ringdown}\end{figure*}

\emph{The ringdown.} A pulse from the generator perturbs the flow on a ring outside the horizon with a pattern of azimuthal number $m$, and the circuit writes, at every tick, the amplitude of that pattern on a second ring (Fig.~\ref{fig:four}, panel 3). For $m=1$ to $6$ the frequencies of the tones on the oscilloscope agree with the board's record and with the numerical twin to $1\%$ or better (Fig.~\ref{fig:ringdown}). All tones decay, so the hole is stable. The longest-lived is the $m=1$ tone, at $0.83\,\OmH$, which rings for nine periods on the oscilloscope. Table~\ref{tab:tones} in the Appendix lists the tones. The absorbing layer that starts at $r=250$ (Appendix) removes short waves but returns part of the long waves of the tones, so the hole rings in a box bounded by the layer: the tones are the quasinormal modes of hole and box together. Moving the inner edge of the layer in the twin between $r=225$ and $290$ leaves the frequencies of $m=2$ to $6$ unchanged within $2\%$ while their decay rates fall steadily with the size of the box: the frequencies are those of the hole, and the decay rates measure the loss per round trip of the box, to the horizon, to the layer and to viscosity. The $m=1$ tone moves with the edge and is a mode of the box. It lies in the superradiant band, $\omega<\OmH$, but its turning radius ($\simeq240$ sites) coincides with the layer, so it reaches the horizon only by tunnelling across the whole barrier; what the rotation returns is outweighed by the loss, a third of the energy per period, and the tone decays. In a twin whose boundary is five times farther the tones give way to ones that decay within one or two periods, the ringdown of the hole alone, which the array of the present board is too small to hold. 

\begin{figure*}[t]\centering
\includegraphics[width=\textwidth]{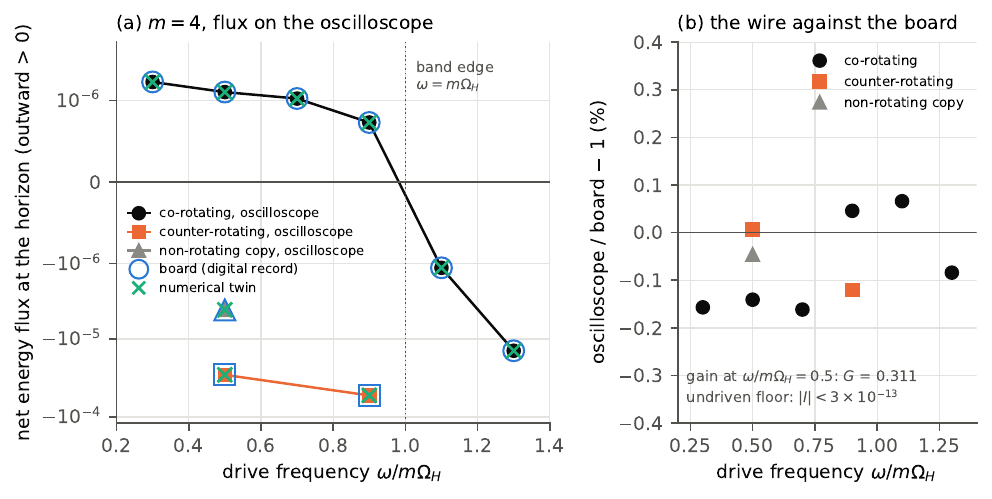}
\caption{Superradiance on the oscilloscope. (a) Net energy flux through the rings two to eight sites outside the
horizon against the drive frequency, for the $m=4$ wave turning with the hole (black), against it (orange),
and on the non-rotating copy (grey); signed logarithmic scale, positive is outward. Filled symbols: oscilloscope;
open circles: board record; crosses: numerical twin. The flux of the co-rotating wave changes sign at the band
edge $\omega=m\OmH$. (b) Oscilloscope against board record, point by point.}
\label{fig:sr}\end{figure*}

\emph{Superradiance.} The two generator channels, in quadrature, drive a wave of azimuthal number $m$ that turns with the hole or against it, and the output is the net energy flux through four rings just outside the horizon (Fig.~\ref{fig:four}, panel 4). The scan on the oscilloscope covers six frequencies of the co-rotating wave, two of the counter-rotating one, the non-rotating copy and the undriven circuit (Fig.~\ref{fig:sr}). The co-rotating flux is outward below $\omega=m\OmH$ and inward above it: the rotating hole gives energy to the wave inside the superradiant band and absorbs it beyond. The counter-rotating wave and the wave on the non-rotating copy are absorbed at every frequency, and the undriven output lies at the resolution of the instrument. Every point agrees with the board's record and with the numerical twin (Fig.~\ref{fig:sr}b); the gain, the ratio of the outward flux to the flux absorbed by the non-rotating copy, is $G=0.311$ on the oscilloscope, the board and the twin alike. This is the gain with the absorbing layer; without it the twin gives $0.41$~\cite{FlachiSW}.

\section*{Conclusions}

We have built an analogue black hole out of logic gates. Its horizon is a property not of the wiring but of the state: a closed curve in a lattice of 32-bit words along which the emergent flow carries disturbances inward exactly as fast as they spread outward. The hole holds that horizon to one part in $10^{4}$ over $10^{5}$ ticks; rings down, in the box set by its absorbing layer, with tones whose frequencies match those of a 64-bit numerical twin to 1\% and whose damping rates match them within the measurement errors; gives energy to co-rotating waves below $\omega = m\Omega_H$ with a gain of $0.31$ and absorbs them above it; and lets a single flipped bit inside the horizon escape only through the dispersive tail of the rule, by an amount that falls exponentially with the depth of the flip. That leak is the signature of the wavenumber-dependent horizon of a medium with superluminal dispersion~\cite{Corley1996,Macher2009}, here measured directly.

Three features set the digital hole apart from its fluid, atomic and optical counterparts. It is exactly deterministic: a run repeats bit for bit, so two copies that differ by a single flipped bit can be subtracted, and the causal structure of the horizon is measured rather than inferred. Its short-distance physics is fixed by the update rule: the scale and the anisotropy of the measured leak agree with the dispersion relation computed from it. And it is reconfigurable: geometry, rotation and dissipation are changed by rewriting static bits, the rule itself by a few hundred lines of logic.

Nothing in the construction is specific to two dimensions or to shallow water. A three-dimensional hole costs only memory and logic. A rule of a few bits per site would let every site carry its own logic and tick at the clock, with intrinsic fluctuations and the throughput to average them; whether a transonic, rotating flow exists in such a rule is an open question. Digital circuits thus form a reconfigurable class of gravitational analogues with the analogue black hole becoming a cellular automaton.

\begin{acknowledgments}
This work was conducted using the Miyabi Supercomputing System at the Joint Center for Advanced High Performance Computing (JCAHPC).\\
\emph{Data availability and AI disclosure.} All data used in this paper are available in the Zenodo repository \href{https://doi.org/10.5281/zenodo.23967575}{10.5281/zenodo.23967575}. AI tools were used to assist with code for the simulation sweeps, data analysis and figure plotting, and with language editing; all figures are plotted directly from the data in the repository. 
\end{acknowledgments}

\clearpage
\onecolumngrid

\section*{Appendix}\label{app:methods}

\subsection*{The rule}
Each site $x$ of a $768\times768$ array holds $25$ single-precision numbers $f_i$, one per velocity $\mathbf e_i=(a,b)$, $a,b\in\{0,\pm1,\pm2\}$, with weights $w_i=w_aw_b$, $w_0=9/16$, $w_{\pm1}=5/24$, $w_{\pm2}=1/96$, and $c_s^2=1/2$. One tick maps $f_i(x,t)$ to $f_i(x,t+1)$:
\begin{enumerate}
\item $h=\sum_if_i$; $\mathbf F=(c_s^2-gh)\nabla h-gh\nabla b+\mathbf F_{\rm s}$, with $g=0.05$, $\nabla$ the centred nearest-neighbour difference, $b(r)$ the bed and $\mathbf F_{\rm s}$ the absorbing layer.
\item $h\mathbf u=\sum_i\mathbf e_if_i+\mathbf F/2$.
\item $f_i^{\rm eq}=h\,w_aw_b\,P_a(u_x)P_b(u_y)$ with $P_0=1-u^2+u^4/2$, $P_{\pm1}=1+u^2-\tfrac56u^4\pm(2u-\tfrac23u^3)$, $P_{\pm2}=1+7u^2+\tfrac{19}6u^4\pm(4u+\tfrac{20}3u^3)$. With $Y_i=f_i-f_i^{\rm eq}+S_i/2$, $S_i$ the forcing term as defined in ref.~\cite{FlachiSW}, the moments of $Y$ relax at rate $0$ (orders $0$, $1$), $1/0.7$ (order $2$; viscosity $c_s^2(\tau-\tfrac12)=0.1$) and $0.6$ (orders $\ge3$).
\item Each $f_i$ moves to $x+\mathbf e_i$.
\end{enumerate}

\subsection*{The hole}
The static word $b_x$ lays out the hole. Beyond $r=300$ (reservoir) each word is reset every tick to the equilibrium of depth $2$ and $\mathbf u=-A/(rh)\,\hat{\mathbf r}+B/r\,\hat{\boldsymbol\theta}$, $A=20$, $B=30$ ($B=0$ for the non-rotating copy). The sill is $b(r)=0.7\,e^{-((r-115)/6)^2}$. The hole is formed with a drain: for $r<105$ the words are multiplied by $1-\min(0.5,\,2\pi A/M)\,w(r)$, with $w$ rising linearly from $0$ at $r=105$ to $1$ at $r=100$ and $M=\sum hw$ over the central sites on the previous tick, so that the volume $2\pi A$ leaves every tick. Once the sill is on the drain is off, and the water leaves through an outflow: inside $r=100$ every word is replaced by the equilibrium of the site at $r=101.5$ on the same ray, captured on the previous tick, with radial velocity raised to at least $1.05\sqrt{gh}$. Neither is local in the sense of Eq.~(\ref{eq:sdc}): the drain uses the sum $M$, fed back by the processor with one tick of lag, and the outflow copies a site up to $100$ sites away, also one tick late. In the held state only the outflow acts; the ring it copies lies $13$ sites inside the horizon, in the supercritical layer. The drive acts on a ring at $170$: $f_i\to f_i[1+a\,e^{-((r-170)/2)^2}\cos(m\theta-\varphi)]$, $a$ and $\varphi$ set every tick from the two inputs. In the ringdown and superradiance runs an absorbing layer $\mathbf F_{\rm s}=-0.02\,s(r)\,h(\mathbf u-\mathbf u_{\rm held})$, $s(r)=\min[1,((r-250)/50)^2]$ beyond $r=250$, damps what reaches it: short waves are absorbed, the long waves of the tones are partly returned. The hole is formed from rest in $32{,}000$ ticks with the drain and without the sill (horizon at $103.2$), the drain is then switched off and the sill and the outflow are switched on for $12{,}000$ ticks, and the state is held for a further $10^5$ ticks ($1.1\times10^5$ ticks after the sill is switched on, Fig.~\ref{fig:hold}); this held state starts every measurement.

\subsection*{The device}
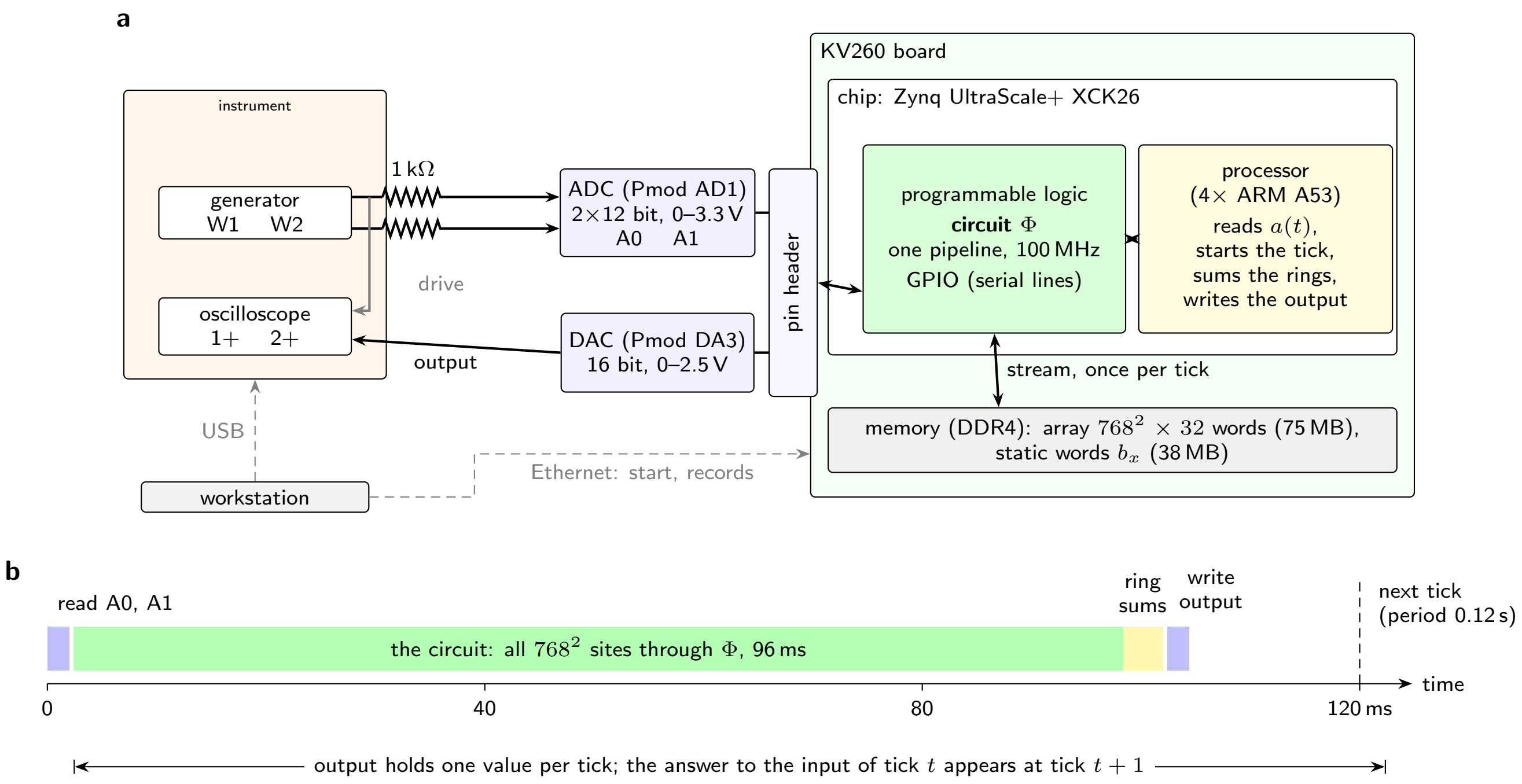
\begin{figure*}[t]\centering
\begin{tikzpicture}[font=\sffamily\scriptsize,>={Stealth[length=1.6mm]},
  box/.style={draw,rounded corners=1.5pt,align=center,inner sep=2.5pt},
  wire/.style={->,thick},
  ctl/.style={->,densely dashed,gray}]
\node[font=\sffamily\bfseries\small,anchor=west] at (-1.1,6.35) {a};
\node[box,minimum width=3.0cm,minimum height=3.3cm,fill=orange!8] (ad3) at (0.6,3.9) {};
\node[anchor=north,align=center,font=\sffamily\tiny] at (ad3.north) {instrument};
\node[box,fill=white,minimum width=2.2cm] (gen) at (0.6,4.15) {generator\\W1 \quad W2};
\node[box,fill=white,minimum width=2.2cm] (osc) at (0.6,2.85) {oscilloscope\\1+ \quad 2+};
\node[box,fill=gray!10,minimum width=2.6cm] (ws) at (0.6,0.9) {workstation};
\draw[ctl] (ws) -- node[left,align=right]{USB} (ad3.south);
\coordinate (w1) at ($(gen.east)+(0,0.18)$); \coordinate (w2) at ($(gen.east)+(0,-0.18)$);
\draw[thick] (w1) -- ++(0.35,0) coordinate(r1a);
\draw[thick,decorate,decoration={zigzag,segment length=1.2mm,amplitude=0.9mm}] (r1a) -- ++(0.7,0) coordinate(r1b) node[midway,above=1.2mm]{1\,k$\Omega$};
\draw[thick] (w2) -- ++(0.35,0) coordinate(r2a);
\draw[thick,decorate,decoration={zigzag,segment length=1.2mm,amplitude=0.9mm}] (r2a) -- ++(0.7,0) coordinate(r2b);
\node[box,fill=blue!6,minimum width=2.2cm,minimum height=1.0cm] (adc) at (5.2,4.15) {ADC (Pmod AD1)\\2$\times$12 bit, 0--3.3\,V\\A0 \quad A1};
\node[box,fill=blue!6,minimum width=2.2cm,minimum height=0.9cm] (dac) at (5.2,2.55) {DAC (Pmod DA3)\\16 bit, 0--2.5\,V};
\draw[wire] (r1b) -- ($(adc.west)+(0,0.18)$); \draw[wire] (r2b) -- ($(adc.west)+(0,-0.18)$);
\draw[wire] (dac.west) -- node[below,pos=0.55]{output} ($(osc.east)+(0,-0.15)$);
\draw[wire,gray] ($(w1)+(0.2,0)$) |- ($(osc.east)+(0,0.18)$);
\node[gray,anchor=west] at (2.35,3.35) {drive};
\node[box,minimum width=6.9cm,minimum height=5.3cm,fill=green!4] (kv) at (10.4,3.55) {};
\node[anchor=north west] at (kv.north west) {KV260 board};
\node[box,minimum width=6.5cm,minimum height=3.15cm,fill=white] (soc) at (10.4,4.1) {};
\node[anchor=north west,align=left] at (soc.north west) {chip: Zynq UltraScale+ XCK26};
\node[box,fill=green!15,minimum width=3.0cm,minimum height=2.15cm,align=center] (pl) at (9.05,3.85) {programmable logic\\[2pt]\textbf{circuit $\Phi$}\\one pipeline, 100\,MHz\\[2pt]GPIO (serial lines)};
\node[box,fill=yellow!15,minimum width=2.9cm,minimum height=2.15cm,align=center] (ps) at (12.15,3.85) {processor\\(4$\times$ ARM A53)\\[2pt]reads $a(t)$,\\starts the tick,\\sums the rings,\\writes the output};
\node[box,fill=gray!12,minimum width=6.5cm,text width=6.3cm,align=center] (ddr) at (10.4,1.55) {memory (DDR4): array $768^2\times32$ words (75\,MB),\\ static words $b_x$ (38\,MB)};
\draw[<->,thick] (pl.south) -- ($(ddr.north)+(-1.3,0)$) node[midway,right]{stream, once per tick};
\draw[<->,thick] (pl.east) -- (ps.west);
\node[box,fill=blue!3,minimum width=0.55cm,minimum height=2.6cm,align=center] (hdr) at (6.75,3.35) {\rotatebox{90}{pin header}};
\draw[thick] (adc.east) -- (adc.east -| hdr.west); \draw[thick] (dac.east) -- (dac.east -| hdr.west);
\draw[<->,thick] (hdr.east) -- ($(pl.west)+(0,-0.6)$);
\draw[ctl] (ws.east) -- ++(1.2,0) |- node[pos=0.75,below]{Ethernet: start, records} ($(kv.south west)+(0,0.5)$);
\end{tikzpicture}

\vspace{4mm}
\begin{tikzpicture}[font=\sffamily\scriptsize,>={Stealth[length=1.6mm]}]
\node[font=\sffamily\bfseries\small,anchor=west] at (-0.6,1.3) {b};
\def\sc{0.125}
\draw[->] (0,0) -- (15.6,0) node[right]{time};
\foreach \t/\l in {0/0,40/40,80/80,120/120\,ms} {\draw (\t*\sc,0) -- ++(0,-0.08) node[below]{\l};}
\fill[blue!25] (0,0.15) rectangle (0.25,0.65); \node[anchor=south west,align=left] at (0.0,0.7) {read A0, A1};
\fill[green!30] (0.3,0.15) rectangle (96*\sc+0.3,0.65);
\node at (48*\sc+0.3,0.4) {the circuit: all $768^2$ sites through $\Phi$, 96\,ms};
\fill[yellow!40] (96*\sc+0.3,0.15) rectangle (96*\sc+0.75,0.65); \node[anchor=south,align=center] at (96*\sc+0.52,0.7) {ring\\sums};
\fill[blue!25] (96*\sc+0.8,0.15) rectangle (96*\sc+1.05,0.65); \node[anchor=south,align=center] at (96*\sc+1.3,0.7) {write\\output};
\draw[densely dashed] (120*\sc,0) -- (120*\sc,1.2); \node[anchor=west,align=left] at (120*\sc+0.1,0.9) {next tick\\(period 0.12\,s)};
\draw[|<->|] (0.3,-0.95) -- (120*\sc+0.3,-0.95) node[midway,fill=white]{output holds one value per tick; the answer to the input of tick $t$ appears at tick $t+1$};
\end{tikzpicture}

\caption{The device. (a)~The bench: generator and oscilloscope, the two converter modules on the pin header, and the board, whose chip holds $\Phi$ in its programmable logic next to a processor; the array lives in the board's memory. Dashed: links used only to start runs and collect records. (b)~One tick, to scale.}
\label{fig:device}\end{figure*}

Figure~\ref{fig:device} shows the signal chain and one tick.

\emph{Signals.} A two-channel generator and oscilloscope (Digilent Analog Discovery~3) form the instrument. The generator channels feed a two-channel $12$-bit ADC (Pmod AD1, $0$--$3.3$~V), sampled once per tick to give $a(t)$; a $16$-bit DAC (Pmod DA3, $0$--$2.5$~V), written once per tick with the observable, feeds oscilloscope channel~1, and channel~2 records one generator channel. Both converters sit on the header of the board, are driven by the processor, and share one ground with the instrument. A workstation programs the instrument over USB and starts runs over Ethernet; neither link carries a signal during a measurement. With no circuit between them, the chain copies the generator to the output within one input step ($0.8$~mV). The board runs at $0.12$~s per tick ($0.2$~s in the ringdown runs; the hold runs unpaced, at about $0.10$~s per tick); one turn of the horizon takes about $2{,}800$ ticks, so every signal lies far below one hertz.

\emph{Board.} On the KV260 board the Zynq UltraScale+ XCK26 combines the programmable logic, which holds $\Phi$, with a processor. The array ($768^2$ sites of $32$ words of $32$ bits, $25$ of them populations, $75$~MB) and the static words ($16$ per site, $38$~MB) are kept in DDR4 and streamed through the logic once per tick; at $100$~MHz and $16.5$ clock cycles per site one pass takes $96$~ms. The device is thus the chip together with the memory and converters of its board.

\emph{One tick.} The processor reads the two inputs, hands $a(t)$ to the logic and starts the tick. The logic passes all $5.9\times10^5$ sites through $\Phi$ in $96$~ms and copies out the flagged rings. From these the processor forms the observable ($z_m$ on a ring, or the energy flux through rings outside the horizon), writes it to the DAC and waits for the end of the period. The output holds one value per tick and answers the input of the previous tick. The processor returns to $\Phi$ only the total depth of the central sites and, in the ringdown runs, the pulse. In the hold and profile runs it also reads the array back every $200$ ticks to form averages over angle; in the ringdown and superradiance runs it never does.

\subsection*{Verification}
The rule is a C++ source synthesised with AMD Vitis and Vivado 2026.1. It was checked against a PyTorch reference at three stages: the source compiled as a program, the synthesised gates in cycle-accurate co-simulation on a $32\times32$ array, and the configured chip read back digitally. All three agree with the reference to the last place of single precision ($2$ to $5\times10^{-7}$ in the $f_i$ after one tick, $8\times10^{-7}$ in the ring captures). Chip and compiled source agree to about one unit in the last place per tick, not bit for bit; the chip repeats itself bit for bit. The circuit closes timing at $100$~MHz and uses $72{,}372$ of the chip's $117{,}000$ look-up tables, $86{,}482$ registers and $252$ of its $1{,}248$ DSP multipliers.

\emph{Numerical twin.} The rule also exists as software: the shallow-water lattice-Boltzmann model of ref.~\cite{FlachiSW}, run in $64$-bit arithmetic on graphics processors. It shares with the device the rule and the layout of the hole, and nothing else. Each measured quantity is available three times: as a voltage on the oscilloscope, as the digital record of the board, and as the value of the twin (Table~\ref{tab:meas}).

\subsection*{Measurement protocols}
Every run begins with $10$ ticks at $0$~V on the output, which marks tick $0$ on the oscilloscope; the board logs what it read and wrote at every tick.

\emph{Hold.} Every $200$ ticks the board reads back the state, averages over angle in rings one site wide, finds by linear interpolation where the inflow speed equals $\sqrt{gh}$, and writes $1.25~\mathrm V+0.1~\mathrm V\times(r_h-113)$.

\emph{Profile.} Every $200$ ticks the processor forms the profile of $\mathrm{Fr}=|u_r|/\sqrt{gh}$ from a read-back; at every tick it reads the probe voltage $V$ on the second input, sets $r=90+70\,V/3~\mathrm V$, and writes $\mathrm{Fr}(r)$ at $1$~V per unit.

\emph{Ringdown.} At a rising edge of the first input above $0.5$~V the processor multiplies the words by $1+8\times10^{-4}e^{-((r-135)/14)^2}\cos m\theta/h$. At every tick the logic captures the ring at $130$ and the processor writes $1.25~\mathrm V+g\,\mathrm{Re}\,z_m$, $z_m=\langle(h-h_{\rm held})e^{-im\theta}\rangle_{\rm ring}$; a second pass writes $\mathrm{Im}\,z_m$.

\emph{Superradiance.} The two inputs, calibrated against the generator, give $c$ and $s$; the drive is $a=2\times10^{-4}\sqrt{c^2+s^2}$, ramped over $4000$ ticks, with $\varphi=\arctan(s/c)$. After $8000$ further ticks, $16{,}000$ ticks are recorded. At every tick the logic captures the rings $r_h+2,+4,+6,+8$; the processor forms $I=2\pi r\langle(g\,\delta h+\mathbf u_{\rm held}\cdot\delta\mathbf u)(h_{\rm held}\,\delta u_r+\delta h\,u_{r,\rm held})\rangle$ on each and writes $1.25~\mathrm V\pm gI_{\rm net}$, the sign alternating from tick to tick.

\subsection*{Error analysis}\label{app:error}
Three effects limit the oscilloscope readings: the resolution of the converters and the noise of the instrument; the offset of the instrument, which is measured and subtracted wherever the output passes through a known level and cancels in the flux; and calibration, of the probe input and of the output chain. Table~\ref{tab:budget} gives the budget.

\subsection*{Converters and noise}
The output step is $38\,\mu$V ($16$ bits over $2.5$~V), the input step $0.8$~mV ($12$ bits over $3.3$~V). The oscilloscope carries $0.5$~mV rms of noise. On the scales used, $1$~mV is $0.01$ sites in the hold, $0.001$ in Fr ($0.006$ sites at the crossing) in the profile, and $1/g$ in the ringdown and flux outputs, $g=2\times10^4$ to $10^6$ being the gain chosen for a swing of about $0.5$~V ($10^9$ for the undriven output).

\subsection*{Offset of the instrument}
The offset wanders by about $1$~mV within an hour and changes from run to run; after a re-zero against levels set by the board the oscilloscope read $1.0003$ times the output plus $4.9\pm1$~mV. Where a run contains a stretch of known output (the ticks before the pulse in the ringdown runs, before the drive in the flux runs) the offset is measured there and subtracted. In the flux runs the sign alternates on alternate ticks, and demodulating with $g\,I_k=s_k\,[y_k-(y_{k-1}+y_{k+1})/2]/2$, $y_k$ the output and $s_k=\pm1$ its sign, removes any offset that varies slowly compared with two ticks.

\subsection*{Calibration}
\emph{Probe input.} At the turning points of the profile sweep, where the one-tick lag does not matter, the radii read by the board and by the oscilloscope agree at the inner end ($90.59$ on both) and differ by $0.22$ sites at the outer end: a $0.3\%$ difference in gain, consistent with the calibration of the input converter ($0.995$). At the horizon the correction is $-0.075$ sites; choosing the turning points from the $0.2$, $0.5$ or $1\%$ extremes of the sweep changes it by about $0.01$.

\emph{Output chain.} Point by point, the flux on the oscilloscope differs from the board's record by $-0.16$ to $+0.07\%$ (mean $-0.07\%$); $0.16\%$ is taken as the calibration uncertainty. It cancels in $G$, a ratio of two fluxes.

\subsection*{Horizon from the profile}
The probe sweeps $90$ to $160$ sites and back twelve times, and the crossing of $\mathrm{Fr}=1$ is found in each sweep by a straight-line fit. The one-tick lag shifts the crossing outward on outward sweeps and inward on inward ones and cancels in their mean; with the probe-gain correction and the uncertainty of the offset the result is $114.35\pm0.03$ sites, against $114.29$ in the board's record.

\subsection*{Tones}
\emph{Box.} The same hole was struck in the same way in twins of $1536^2$ and $3072^2$ words, with the free disc extended to $r=750$ and $1500$ and the damping layer moved to its edge. The horizon radius and $\OmH$ are unchanged to $0.3\%$, and none of the tones of Table~\ref{tab:tones} appears; the pencil finds instead poles at $\omega=1.2$ to $1.6\,m\OmH$ decaying at $0.4$ to $0.8\,m\OmH$, within one to two periods, with amplitudes ten times smaller. In the $3072^2$ twin these poles are stable at every $m\ge2$ once the undriven response of the lattice, recorded without a pulse and at $m=4$ as large as the pulse's, is subtracted, and they are unchanged when the pulse is doubled; the $1536^2$ twin agrees where its extraction is stable ($m=5$). \emph{Radius of the layer.} In the $768^2$ twin the absorbing layer, which starts at $r=250$ in the table, was moved to $150$, $200$, $225$ and $290$. From $225$ to $290$ the frequencies of $m=2$ to $6$ stay within $2\%$ of the table while the decay rates fall monotonically with the radius, by a factor of $1.3$ to $2.5$; at $150$ the tones are replaced by the damped family of the large boxes. The $m=1$ frequency follows the radius ($0.99$, $0.92$, $0.84$, $0.72\,\OmH$ at $200$, $225$, $250$, $290$): it is a mode of the box. The tones of the table are therefore those of the hole in the box bounded by the layer, the frequencies set by the hole and the decay rates by the box. Each tone is extracted from ticks $1200$ to $3900$ after the pulse by a matrix pencil at model orders $4$ to $10$, keeping poles that recur at five or more orders; the same extraction is applied to the oscilloscope, the board's record and the numerical twin, all on ring $130$. The error combines the scatter of the pole over the orders with the difference between oscilloscope and board record, which measures how instrument noise moves the extraction. The $m=1$ tone ($30{,}600$ ticks) is extracted from the in-phase pass alone by the cluster method, the scatter over windows being the error. The offset is taken from the ticks before the pulse, and poles with $\mathrm{Re}\,\omega<10^{-3}$ per tick, which describe its drift, are excluded. On the wire the frequencies agree with the numerical twin within $0.012\,m\OmH$ and the damping rates within $0.016\,m\OmH$ (Table~\ref{tab:tones}).

\subsection*{Energy flux and gain}
Each run lasts $28{,}000$ ticks with chopped output. The flux is the mean of the demodulated ticks after the transient; the statistical error is the scatter of the means of four equal windows divided by two, $0.005$ to $0.17\%$. Without drive the output is $(-7\pm2)\times10^{-14}$ on the oscilloscope and $-2\times10^{-13}$ in the board's record, the resolution at the gain used ($10^9$). The gain $G=I_{\rm co}/|I_{\rm nr}|$, the co-rotating flux over that of the non-rotating copy, at $\omega/m\OmH=0.5$ is $0.3107\pm0.0002$ on the oscilloscope and $0.3110$ in the board's record. The error is statistical. The flux is outward in the band with every treatment of the outer boundary; the size of the gain, a ratio to the flux absorbed by the non-rotating copy, varies with that treatment between $0.31$ and $0.41$ over the protocols of ref.~\cite{FlachiSW}.

\subsection*{The causal cone and the dispersion of the rule}\label{app:cone}
\emph{Protocol.} From the held state the chip runs $400$ ticks; every two ticks the board reads back, for a $401\times401$ window around the centre, the depth $h$ and a $32$-bit hash of the $25$ words of each site. The run is repeated with bit $20$ of $f_0$ flipped at $t=0$ at the site nearest a chosen radius ($8\%$ of that number), keeping per radial bin of one site the sum of $\delta h^2$, the largest $|\delta h|$ and the number of sites whose hashes differ, and the full map of $\delta h$ at $13$ ticks. A third run without the flip reproduced the first at every site and tick. On the chip the flips were placed on one ray at $11$ radii from $104.5$ to $124.5$; on the software circuit at $15$ radii and three angles ($0^\circ$, $22.5^\circ$, $45^\circ$). The level of rounding is set by a flip of bit $13$, whose effect is lost after about $60$ ticks. The outward speed is that of the centroid of the energy of the difference beyond the launch radius, followed over $30$ ticks.

\begin{figure*}[t]\centering
\includegraphics[width=0.95\textwidth]{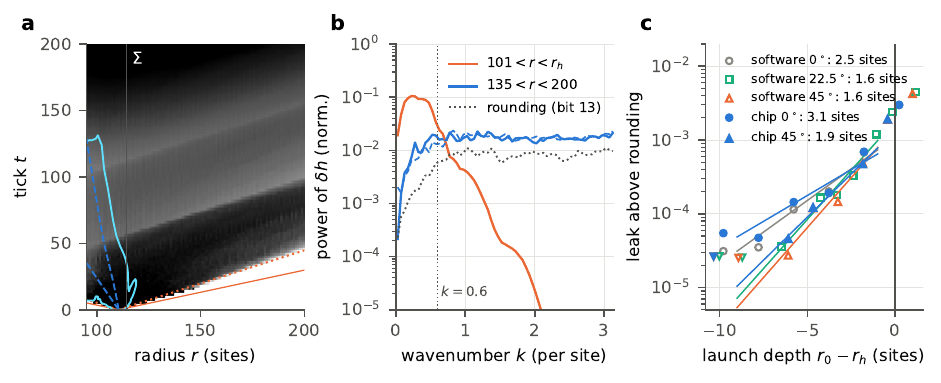}
\caption{Post-processing of the causal-cone read-backs. (a)~Fraction of the sites within $2t$ of the flipped bit ($r_0=110.5$, chip) whose words differ, against radius and tick (grey scale, $0$ to $1$); dotted, $r_0\pm2t$; thin, $r_0\pm3t$, the reach of the push; dashed, the characteristics of the fluid limit; cyan, the contour at $10^{-4}$ of the energy of the difference, the cone of Fig.~\ref{fig:cone}b. (b)~Power spectrum of $\delta h$ at $t=40$ for a flip at $r_0=112.5$, inside the horizon ($101<r<r_h$) and beyond the detection circle ($135<r<200$), with the rounding floor (flip of bit $13$) in the outer annulus; solid, chip; dashed, software circuit. (c)~Leak above rounding against depth for the three launch angles of the software circuit and for the chip at $0^\circ$ and $45^\circ$, with the fitted decay lengths.}
\label{fig:cone2}\end{figure*}

\begin{figure*}[t]\centering
\includegraphics[width=0.95\textwidth]{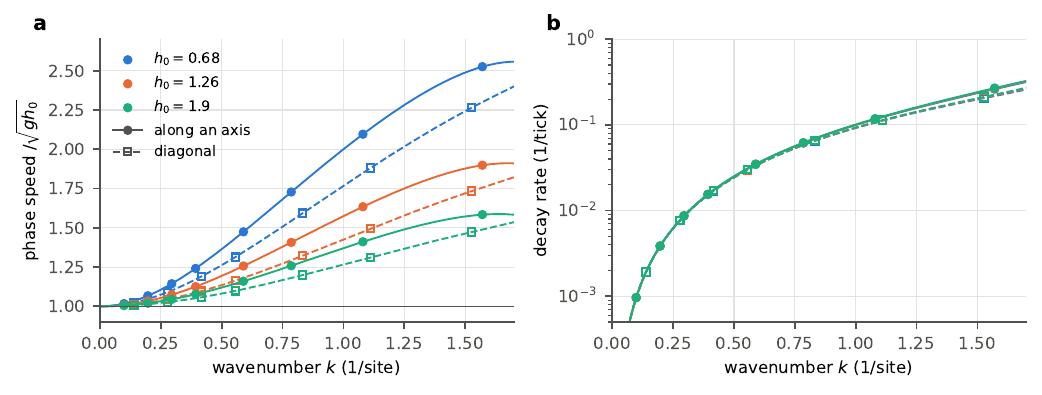}
\caption{Dispersion of the rule on a still layer. (a) Phase speed over $\sqrt{gh_0}$ against wavenumber along an axis (filled) and the diagonal (open), at the depths of the horizon ($0.68$), of $120$ sites ($1.26$) and of $150$ sites ($1.9$); points, the rule run in 64-bit arithmetic; lines, the exact dispersion of the linearised rule (solid, axis; dashed, diagonal). (b) Decay rate per tick. Above $k\approx0.6$ the waves decay within a period.}
\label{fig:disp}\end{figure*}

\emph{Wiring cone, wave-number content and anisotropy.} Figure~\ref{fig:cone2} shows three further quantities formed from the same read-backs. (a)~The number of sites per radius at which any of the $25$ words differs, from the hashes, fills the square reached by the wiring: within $2t$ of the flipped bit over $90\%$ of the words differ after $20$ ticks, while the energy of the difference stays within the sound cone. The words differ out to $3t$ for the first ticks, because the push reads the depth at the nearest neighbours before the populations move by up to two sites, so that the reach of $\Phi$ is $R=3$; the differences carried by the push alone fall below rounding within about $40$ ticks, after which the set grows at two sites per tick. The sites $r<105$ are left out: inside $r=100$ the words are copies of the ring at $101.5$ (the outflow), so a difference that reaches that ring appears along the whole ray inside it on the next tick. (b)~Inside the horizon $90\%$ of the power of $\delta h$ lies at $k<0.6$; beyond $135$ sites, at $t=40$, the rounding floor accounts for $46\%$ of the power and $83\%$ of the excess lies at $k>0.6$, where the waves of Fig.~\ref{fig:disp} outrun the inflow. (c)~The decay length of the leak, fitted as in Fig.~\ref{fig:cone}e for each launch angle, is $2.5\pm0.3$ sites at $0^\circ$, $1.6\pm0.2$ at $22.5^\circ$ and $1.6\pm0.3$ at $45^\circ$ on the software circuit ($2.3\pm0.3$ pooled, the fit drawn in Fig.~\ref{fig:cone}e), and $3.1\pm0.6$ on the chip at $0^\circ$ and $1.9\pm0.2$ at $45^\circ$ (seven flips at radii $104$ to $120$, run after a clean restart of the board; at this angle the chip and the software circuit give the same escaped energy against depth to $1$--$3\%$). The leak does not depend on the kick: on the software circuit, flipping bit $18$, $20$ or $22$ of $f_0$, or bit $22$ of an outward or inward population ($f_5$, $f_{10}$) instead, gives the same escaped energy against depth and decay lengths of $2.3\pm0.4$, $2.1\pm0.5$ and $1.5\pm0.2$ sites. Nor does it depend on the measure of energy: with the momentum read back as well, the second-order wave energy $g\,\delta h^2/2+h_0|\delta\mathbf u|^2/2+\delta h\,\mathbf u_0\!\cdot\!\delta\mathbf u$ of the difference is about a tenth of $\sum\delta h^2$ at launch, the rest being vortical, and the outward flux of this energy through the circle at $135$ sites, accumulated in time, saturates within $100$--$170$ ticks and stays flat to $400$: its final value falls off with depth over $2.1\pm0.3$ sites against $2.5\pm0.3$ for the snapshot of Fig.~\ref{fig:cone}e on the same runs. The cone runs have no absorbing layer; the reservoir at $300$ sites is the outer boundary, too far for anything to return within $400$ ticks, and the circle at $135$ sites lies outside the ergosurface at $121$.

\emph{Dispersion and the leak.} Linearising one tick about a uniform layer of depth $h_0$ and velocity $\mathbf u_0$ gives, for each wavevector $\mathbf k$, a $25\times25$ matrix whose eigenvalues $e^{-i\omega-\gamma}$ hold the frequencies and decay rates of the modes; two of them are the sound waves (lines in Fig.~\ref{fig:disp}, for a still layer). Runs of the rule in 64-bit arithmetic on a still layer in a periodic $64\times64$ box (points) follow the linearised rule to $2\times10^{-10}$ of the perturbation. Long waves travel at $\sqrt{gh_0}$, shorter ones faster, to leading order as $c^2(k)=gh_0\,\sin k/k+c_s^2(1-\sin k/k)$: the hydrostatic force replaces $c_s^2h$ by $gh^2/2$ only for disturbances long compared with the two-site difference that computes it. Waves shorter than about ten sites ($k\gtrsim0.6$) travel at $0.3$--$0.5$ sites per tick, up to $2.5$ times $\sqrt{gh_0}$ at the depth of the horizon, and decay within $4$--$30$ ticks.

The fall-off of the leak with depth follows from this dispersion, with nothing fitted. On the held flow the linearised rule gives the frequency $\omega(k,r)$ and the decay rate $\gamma(k,r)$ of the outgoing sound wave along a ray. Inside $\Sigma$ the long waves have $\omega<0$ and are carried inward; a component launched at depth $d$ climbs out only if its phase speed exceeds the inflow there, that is above a threshold $k_{\min}(d)$, which rises from $0.4$ at $d=2$ to $0.7$ at $d=8$ sites along an axis, and from $0.5$ to $0.9$ along a diagonal. As it climbs its frequency is conserved, its wavenumber falls and it decays at the rate $\gamma$, so the escaping fraction of the flat spectrum of a flipped bit is $\int_{k>k_{\min}}dk\,\exp[-2\int\gamma\,dr/v_g]$, with $v_g=\partial\omega/\partial k$ and the inner integral taken along the ray to $135$ sites. The short waves found outside at tick $40$ (Fig.~\ref{fig:cone2}b) are the faster and more strongly damped part, which arrives first. The computed leak falls off over $1.9$ to $2.2$ sites along an axis and $1.4$ to $1.6$ along a diagonal, for spectra flat in one and in two dimensions. The measured lengths are $2.5\pm0.3$ and $1.6\pm0.3$ on the software circuit and $3.1\pm0.6$ and $1.9\pm0.2$ on the chip: the computation gives the scale of the leak and its dependence on direction, with lengths shorter than the measured ones by up to about two standard errors. From a depth of eight sites a flipped bit delivers about $5\times10^{-3}$ of what it delivers from outside, against at most $10^{-4}$ for smooth depth bumps $1.5$ to $3$ sites wide~\cite{FlachiSW}, which excite much less of the short waves. The leak responds only weakly to the relaxation rate of the moments of order $\ge3$: varying it from $0.3$ to $1.4$ (the published value is $0.6$) moves the horizon by $0.04$ sites and the decay length from $2.7\pm0.5$ to $1.8\pm0.3$ sites, because the damping at $k\lesssim1$ is set by the viscosity.

\clearpage

\begin{table*}[t]\centering\small
\begin{tabular}{@{}lccc@{}}
\toprule
quantity & oscilloscope & board record & numerical twin \\
\midrule
horizon radius, rotating hole (sites) & $114.28\pm0.01^{+0.01}_{-0.10}$ & $114.29$ & $114.27$ \\
horizon radius, non-rotating copy (sites) & $114.80\pm0.01^{+0.01}_{-0.10}$ & $114.72$ & $114.72$ \\
horizon from the flow profile (sites) & $114.35\pm0.03$ & $114.29$ & $114.27$ \\
tones, $m=1$ to $6$ (frequency) & \multicolumn{3}{c}{agree to $1\%$ or better (Table~\ref{tab:tones})} \\
energy flux, co-rotating, $\omega/m\OmH=0.5$ & $+1.3027(8)\times10^{-6}$ & $+1.305\times10^{-6}$ & $+1.302\times10^{-6}$ \\
energy flux, non-rotating copy, $\omega/m\OmH=0.5$ & $-4.1924(5)\times10^{-6}$ & $-4.194\times10^{-6}$ & $-4.186\times10^{-6}$ \\
gain $G$ at $\omega/m\OmH=0.5$ & $0.3107(2)$ & $0.3110$ & $0.311$ \\
sign change of the co-rotating flux & \multicolumn{3}{c}{between $\omega/m\OmH=0.9$ and $1.1$} \\
\bottomrule
\end{tabular}
\caption{The same quantities on the oscilloscope, in the digital record of the board, and in the numerical twin. Oscilloscope values carry their statistical error and, for the held radii, the systematic of the instrument offset (Table~\ref{tab:budget}); fluxes in the last digits carry the statistical error only, with a calibration uncertainty of $0.16\%$.}
\label{tab:meas}\end{table*}

\begin{table*}[t]\centering\small
\begin{tabular}{@{}lccccc@{}}
\toprule
source & size & horizon hold & horizon profile & tones & energy flux \\
\midrule
output converter, $16$ bit & $38\,\mu$V & $0.0004$ & $<0.001$ & -- & $<10^{-5}$ \\
oscilloscope noise & $0.5$~mV rms & $0.008$ & $0.003$ & in extraction & $\le2\times10^{-3}$ \\
input converter, $12$ bit & $0.8$~mV & -- & $<0.003$ & -- & -- \\
offset of the instrument & $-1$ to $+10$~mV & $^{+0.01}_{-0.10}$ & $0.03$ & removed & cancels \\
probe gain & $0.4\%$ & -- & $0.075\pm0.01$ (corrected) & -- & -- \\
output gain & $\le0.16\%$ & -- & -- & -- & $1.6\times10^{-3}$ \\
lag of one tick & $0.12$~s & -- & $\pm0.027$ (cancels) & -- & -- \\
extraction & & -- & -- & $0.001$--$0.018$ & -- \\
\bottomrule
\end{tabular}
\caption{Uncertainty budget of the oscilloscope readings. Horizon radii in sites; tones in units of $m\OmH$; energy flux as a relative error. ``Corrected'': a measured shift that was removed; the number is what remains.}
\label{tab:budget}\end{table*}

\begin{table*}[t]\centering\small
\begin{tabular}{@{}ccccccc@{}}
\toprule
 & \multicolumn{2}{c}{oscilloscope} & \multicolumn{2}{c}{board record} & \multicolumn{2}{c}{numerical twin} \\
\cmidrule(lr){2-3}\cmidrule(lr){4-5}\cmidrule(lr){6-7}
$m$ & $\mathrm{Re}\,\omega/m\OmH$ & $-\mathrm{Im}\,\omega/m\OmH$ & Re & $-$Im & Re & $-$Im \\
\midrule
1 & $0.835(2)$ & $0.028(1)$ & $0.834$ & $0.029$ & $0.840$ & $0.027$ \\
2 & $1.958(9)$ & $0.229(9)$ & $1.955$ & $0.229$ & $1.962$ & $0.230$ \\
3 & $1.590(5)$ & $0.137(9)$ & $1.596$ & $0.128$ & $1.602$ & $0.132$ \\
4 & $1.439(13)$ & $0.095(10)$ & $1.426$ & $0.098$ & $1.429$ & $0.100$ \\
5 & $1.695(1)$ & $0.134(4)$ & $1.695$ & $0.134$ & $1.700$ & $0.135$ \\
6 & $1.550(13)$ & $0.128(18)$ & $1.561$ & $0.111$ & $1.561$ & $0.112$ \\
\bottomrule
\end{tabular}
\caption{Tones of the ringdown on ring $130$, frequency and damping rate in units of $m\OmH$: oscilloscope (both quadratures, from two passes; $m=1$ from the in-phase pass alone), the board's digital record of the same runs, and the numerical twin, all with the same extraction. Errors in the last digits. The tones are those of the hole in the box bounded by the absorbing layer at $250$, the frequencies set by the hole and the decay rates by the box; in twins with the boundary at $750$ and $1500$ they are absent (Appendix, Tones).}
\label{tab:tones}\end{table*}

\clearpage
\twocolumngrid

\end{document}